# Extreme Scaling of Lattice Quantum Chromodynamics

David BRAYFORD, Momme ALLALEN and Volker WEINBERG
*Leibniz-Rechenzentrum der Bayerischen Akademie der Wissenschaften*

**Abstract.** As the complexity and size of challenges in science and engineering are continually increasing, it is highly important that applications are able to scale strongly to very large numbers of cores (>100,000 cores) to enable HPC systems to be utilised efficiently. This paper presents results of strong scaling tests performed with an MPI only and a hybrid MPI + OpenMP version of the Lattice QCD application BQCD on the European Tier-0 system SuperMUC at LRZ.

## Introduction

Within the standard model of elementary particle physics quantum chromodynamics (QCD) describes the strong interactions between quarks and gluons, the smallest building blocks of matter. The equations of this theory are so complicated that they cannot be solved by traditional perturbative methods of quantum field theory. The only computational *ab initio* approach for solving QCD is Lattice QCD. In order to simulate QCD on HPC systems, one approximates the space-time continuum by a four dimensional finite box divided into a discrete set of points, the lattice. Due to continuous algorithmic improvements and the advent of Petaflops-scale computing facilities, lattice QCD simulations have reached a level where realistic simulations with physical quark masses have become possible and allow for a first principle study of strongly interacting elementary particles.

BQCD (Berlin Quantum ChromoDynamics) is a Hybrid Monte-Carlo (HMC) code written in Fortran that simulates QCD with dynamical Wilson fermions. Beyond being widely used in the lattice QCD community, BQCD is also included in the Unified European Applications Benchmark Suite (UEABS) of PRACE, the Partnership for Advanced Computing in Europe. The kernel of the program is a standard conjugate gradient solver with even/odd pre-conditioning. In a typical HMC run between 80% and 95% of the total computing time is used for the multiplication of a very large sparse matrix ("hopping matrix") with a vector. At the single CPU level QCD programs benefit from the fact that the basic operation is the multiplication of small complex matrices.

QCD programs are parallelised by decomposing the lattice into regular domains. The nearest neighbour structure of the hopping matrix implies that the boundary values (surfaces) of the input vector have to be exchanged between neighbouring processes in every iteration of the solver. The boundary exchange is communication intensive because the local lattices are typically small and the surface to volume ratio is quite large. BQCD has various communication methods implemented in the hopping matrix multiplication:

MPI, OpenMP, a hybrid combination of both, as well as shmem (single sided communication).

**1. Execution Environment**

The initial dynamical lattice QCD computations at LRZ have become possible with the first national supercomputer at LRZ, the Hitachi SR8000-F1 machine HLRB-I, installed in 2000 [1, 2]. Since then BQCD has been ported to several machines and architectures [3, 4]. On the next national supercomputer at LRZ, the SGI 4700 Altix HLRB-II, BQCD has shown good scaling on up to 8K cores for both the pure MPI and the hybrid version of the code.

In this paper we discuss the scaling behaviour of BQCD on the current 3 Pflop/s SuperMUC system at LRZ. SuperMUC [5] serves as a Tier-0 system within PRACE. It is an IBM System x iDataPlex machine based on Intel processors and Infiniband technology. SuperMUC consists of 18 thin node islands equipped with Intel SandyBridge processors and one fat node island with Intel Westmere-EX processors. The thin node islands are connected via a fully non-blocking Mellanox FDR-10 Infiniband network. Each thin node island consists of 512 nodes of 8 SandyBridge-EP Intel Xeon E5-2680 processors, providing 8192 cores per island.

On SuperMUC we have investigated the scaling of an MPI only and a hybrid MPI + OpenMP version of the code. BQCD was built with the Intel ifort 12.1 Fortran compiler and Intel MPI 4.1. All scaling tests have been performed on the thin node islands of SuperMUC.

Scaling has been studied for two lattices sizes, $96^3 \times 192$ and $64^3 \times 96$, for the MPI only and the hybrid MPI + OpenMP version, respectively. The lattice sizes have been chosen in order to fit the domain decomposition and the model parameters with respect to the memory size of the system. Similar lattice sizes have been analysed on the Blue Gene/P machine during the Extreme Scaling Workshop 2010 in Jülich [4].

## 2. Results

### 2.1. MPI only Version of BQCD for a $96^3 \times 192$ Lattice

Performance results for the MPI only version of the code are summarised in Table 1. Figure 1 shows the strong scaling of BQCD on up to 16K cores for a lattice size of $96^3 \times 192$. Unfortunately, we were unable to acquire data for 32K, 64K and 128K cores, because the application was either crashing or running extremely slowly for still unknown reasons[1]. Within one thin node island of SuperMUC, i.e. for up to 8K cores, scaling of BQCD is almost linear. For 16K cores (2 full islands) the overall performance significantly drops below linear.

**Table 1.** Lattice decomposition, total time, mean performance per core and overall performance of the conjugate gradient solver of BQCD for a $96^3 \times 192$ lattice using the MPI only version.

| # Cores | Local Lattice | Total Time [s] | Mean Perf. per Core [Mflop/s] | Overall Perf. [Gflop/s] |
|---|---|---|---|---|
| 1024 | $96 \times 12 \times 12 \times 12$ | 8096.74 | 1821.44 | 1865.15 |
| 2048 | $48 \times 12 \times 12 \times 12$ | 4403.19 | 1674.66 | 3429.15 |
| 4096 | $24 \times 24 \times 12 \times 6$ | 2762.66 | 1334.56 | 5466.35 |
| 8192 | $24 \times 12 \times 12 \times 6$ | 1661.58 | 1109.46 | 9088.72 |
| 16384 | $12 \times 6 \times 12 \times 12$ | 1298.62 | 709.78 | 11628.97 |

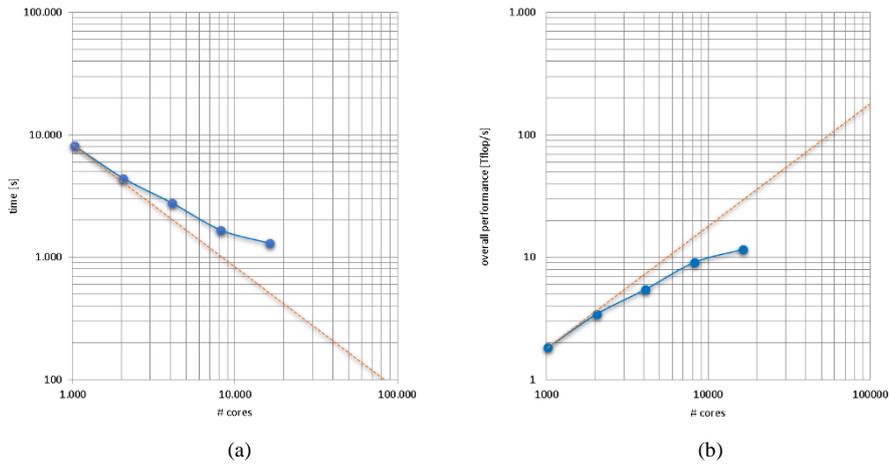

(a)      (b)

**Figure 1.** Strong scaling of the conjugate gradient solver of BQCD using the MPI only version of the code for a $96^3 \times 192$ lattice. Subfigure (a) shows the total time spent within the solver and (b) the overall performance as a function of the total number of cores. The straight dotted line indicates linear scaling.

---

[1] For example, the 32K job could not complete after an execution time which was four times longer than the time the 16K job needed to complete. This seems to be an Intel MPI issue and not a problem of the BQCD code.

## 2.2. Hybrid MPI + OpenMP Version of BQCD for a $64^3 \times 96$ Lattice

Performance results for the hybrid MPI + OpenMP version of the code are summarised in Table 2. Each MPI task executes 8 OpenMP threads. The plots in Figure 2 show that the hybrid version scales well within 2 islands (16384 cores). Negative scaling is observed from 64K to 128K cores. The performance decreases observed between 64K and 128K cores with the hybrid version of BQCD is due to the small local lattice sizes. The data from a relatively large surface of the small domains has to be communicated to the neighbour processes and challenges the communication network for large job sizes.

**Table 2.** Lattice decomposition, total time, mean performance per core and overall performance of the conjugate gradient solver of BQCD for a $64^3 \times 96$ lattice using the hybrid MPI + OpenMP version.

| # Cores | Local Lattice | Total Time [s] | Mean Perf. per Core [Mflop/s] | Overall Perf. [Gflop/s] |
|---|---|---|---|---|
| 1024*8= 8192 | $16 \times 16 \times 8 \times 12$ | 29.62 | 930.09 | 7619.30 |
| 2048*8= 16384 | $16 \times 16 \times 8 \times 6$ | 17.08 | 806.47 | 13213.21 |
| 8192*8= 65536 | $16 \times 8 \times 4 \times 6$ | 8.10 | 424.96 | 27849.92 |
| 16384*8= 131072 | $8 \times 8 \times 4 \times 6$ | 10.45 | 164.73 | 21591.83 |

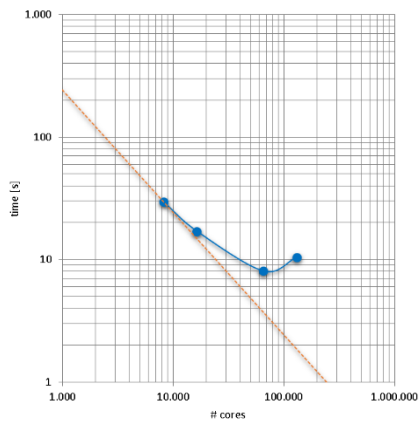
(a)

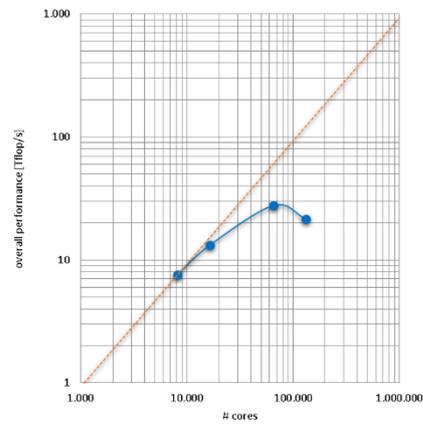
(b)

**Figure 2.** Strong scaling of the conjugate gradient solver of BQCD using the hybrid MPI + OpenMP version of the code for a $64^3 \times 96$ lattice. 8 OpenMP threads are executed per MPI task. Subfigure (a) shows the total time spent within the solver and (b) the overall performance as a function of the total number of cores. The dotted straight line indicates linear scaling

## 3. Conclusion

We have shown that BQCD scales well on the SuperMUC machine at LRZ with both the MPI only and the hybrid MPI + OpenMP versions of the code within one (8k cores) or two (16k cores) thin node islands, respectively. The scaling properties of BQCD strongly depend on the local lattice decomposition. The optimal values are highly dependent upon the architecture of the HPC system used. It is important to tune the local lattice sizes to achieve optimal scaling performance and ensure efficient communication between the processors. It seems that there are some underlying issues, which result in BQCD crashing or running extremely slowly with very large number of MPI jobs. This is still not clearly understood yet and requires further analysis. To be able to better compare results obtained with the two versions of the code with each other and also with results from other HPC systems, further runs at various lattice sizes will be done in future. Also more investigation on the weak scaling behaviour of BQCD is needed on SuperMUC.

## Acknowledgements

Our work was financially supported by the KONWIHR project OMI4papps ("Optimisation, Modelling and Implementation for highly parallel applications").

## References

[1] G. Schierholz, *Simulation of QCD with Dynamical Quarks*, in: S. Wagner, W. Hanke, A. Bode and F. Durst (Eds.), *High Performance Computing in Science and Engineering*, Munich 2002, Springer-Verlag, 367-377.
[2] G. Schierholz and H. Stüben, *Optimizing the Hybrid Monte Carlo Algorithm on the Hitachi SR8000*, in: S. Wagner, W. Hanke, A. Bode and F. Durst (Eds.), *High Performance Computing in Science and Engineering*, Munich 2004, Springer-Verlag, 385-393.
[3] M. Allalen, M. Brehm and H. Stüben, *Performance of Quantum Chromodynamics (QCD) Simulations on the SGI Altix 4700*, *Computational Methods in Science and Engineering*, 14(2), 69-75 (2008).
[4] H. Stüben and M. Allalen, *Extreme Scaling of the BQCD Benchmark*, in: B. Mohr and W. Frings (Eds.), Technical Report FZJ-JSC-IB-2010-03 (2010) 31–34.
[5] http://www.lrz.de/services/compute/supermuc/systemdescription/